\documentclass[12pt]{iopart}
\usepackage{color}
\usepackage{graphics}
\usepackage{epsfig}
\usepackage{float}
\begin{document}

\title{Precursor magneto-sonic solitons in a plasma from a moving charged object}

\author{Atul Kumar and Abhijit Sen}

\address{Institute for Plasma Research, Bhat , Gandhinagar - 382428, India}
\ead{atul.j1211@gmail.com, and senabhijit@gmail.com}
\vspace{10pt}

\begin{abstract}
The nature of fore-wake excitations created by a charged object moving in a magnetized plasma is investigated using particle-in-cell simulations. Our studies establish for the first time the existence of precursor magneto-sonic solitons traveling ahead of a moving charged object. The nature of these excitations and the conditions governing their existence are delineated. We also confirm earlier molecular dynamic and fluid simulation results related to electrostatic precursor solitons obtained in the absence of a magnetic field. The electromagnetic precursors could have interesting practical applications such as in the interpretation of observed nonlinear structures during the interaction of the solar wind with the earth and the moon and may also serve as useful tracking signatures of charged space debris traveling in the ionosphere. 
\end{abstract}

%
%
%
%
%

\section{Introduction}
\label{intro}
Precursors that travel faster than a source pulse exciting them are of much fundamental and practical interest in many areas of physics. For example, optical precursors that were theoretically predicted many years ago by Sommerfeld and Brillouin \cite{Sommerfeld1914,Brillouin1914}, provide insights into questions related to group velocity in a dispersive medium \cite{Brillouin1960} and can be potentially useful in applications like underwater communications \cite{Choi2004} or deep imaging of biological tissue \cite{Albanese1989}. Optical precursors have been experimentally observed when light pulses have been launched in a deionized water column \cite{Choi2004} and also in a linear dispersive medium consisting of  gas of cold potassium atoms \cite{Jeong2006}. Their occurrence is associated with the existence of anomalous dispersion in the medium, namely, a negative value of the derivative of the refractive index with respect to frequency at the source pulse carrier frequency. In principle, such a linear phenomenon can also take place in a gaseous plasma medium as has been pointed out in past studies \cite{Pleshko1969,Buckley1979,Cartwright2009}. However, in a plasma there can be other competing mechanisms of precursor generation that do not depend on anomalous dispersion but exploit the nonlinear properties of the medium. One such process is the excitation of nonlinear structures like solitons or shocks that can propagate faster than the moving excitation source. The generation of such nonlinear precursor pulses has been conjectured a long time ago in hydrodynamics and studied in the context of fore-wake structures generated by moving ships \cite{Ertekin1984,Akylas1984,Lee1989} but has received scant attention in plasmas until recently. Theoretical studies in the framework of a fluid model have now shown the existence of  electrostatic precursor pulses ahead of a super-sonically moving charged object in a plasma \cite{Sen2015,Sanat_prec,Sanat_md2016}, and experimental support has come from  observations in a laboratory dusty plasma device \cite{Surabhi2016,Garima2019}. One of the advantages of these solitonic precursors over the Brillouin or Sommerfeld type precursors is that they can propagate over long distances without any attenuation and can thus prove more useful in communication applications or in transferring packets of energy to a target. To date, the theoretical/simulation studies of such precursors in plasmas have been limited to the excitation of low frequency electrostatic pulses in a fluid plasma model \cite{Sen2015,Sanat_prec} or in a strongly coupled Yukawa fluid system \cite{Sanat_md2016}.  In the present work, we take a more fundamental approach of investigating this phenomenon at a particulate level and  consider the excitation of electromagnetic pulses.  Using two dimensional particle-in-cell simulations we establish from first principles the existence of precursor magnetosonic solitons propagating ahead of a moving charged object in a magnetized plasma and explore the excitation conditions and the nature of these solitons. We also point out the potential uses of our results in a few practical 
applications ranging from space debris detection to interpretation of precursor waves in earth solar wind interactions. \\

The paper is organized as follows. In section \ref{model} we briefly describe the physical model used for our simulations and give the essential computational details. Our main results are presented in section \ref{results} with separate subsections devoted to confirmation of earlier electrostatic precursor results (\ref{ESP}), magnetosonic solitons
(\ref{EMP}), pinned solitons (\ref{PIN}) and precursors from a two dimensional source (\ref{2d}). Section \ref{disc} 
gives a summary of our results and discusses their possible occurrences in nature and their potential applications. 

\section{Simulation Model}
\label{model} 
\begin{figure}[h]
\center
                \includegraphics[width=0.5\textwidth]{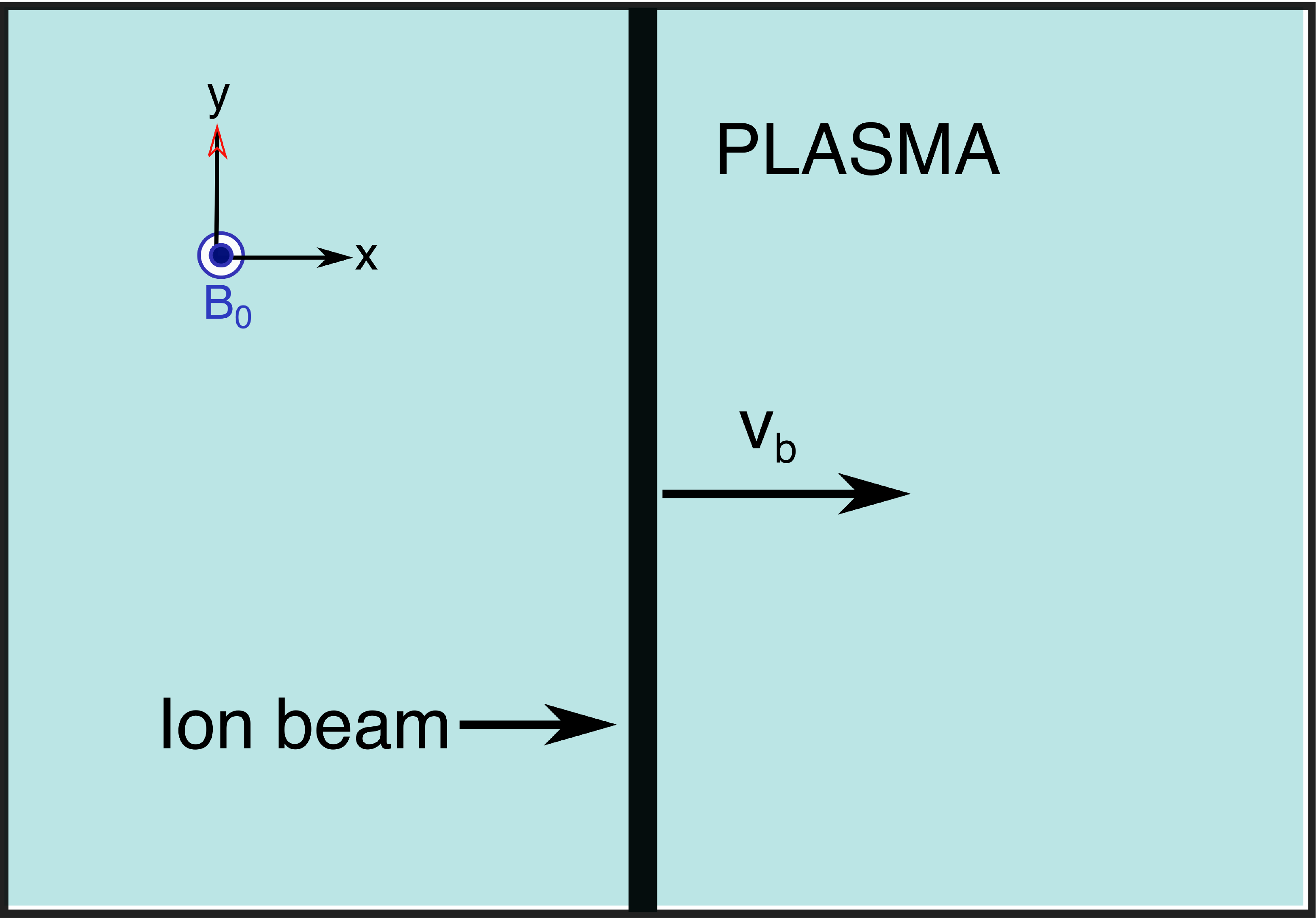} 
            \caption{Schematic diagram (not to scale) of the simulation geometry where an ion beam (thick black line) representing  an idealized finite charge source is moving with a speed $V_b$ in the positive $\hat{x}$-direction. A magnetic field is applied in the positive $\hat{z}$ direction.}  
                 \label{fig1}
                 
         \end{figure}

\noindent
Our 2d-3V particle-in-cell simulations have been carried out using the OSIRIS-4.0 code \cite{hemker,Fonseca2002,osiris} for a quasi-neutral plasma system consisting of positive ions and electrons. For simplicity and shorter simulation times the ion to electron mass ratio has been taken to be $25$. In this system, a short pulsed beam of positive (or negative) ions is made to travel at various velocities and the resultant wave excitations are studied. 
A constant magnetic field $B_0$ is imposed in the $\hat{z}$ direction while the beam travels in the 
$\hat{x}$ direction. A schematic of the simulation geometry and the start-up scenario is shown in Fig.~\ref{fig1}.
The magnetic field strength is chosen to be such that the following condition holds, namely, $ \omega_{ci} < \omega_{pi} \ll \omega_{pe} < \omega_{ce}$, where $\omega_{ci}$, $\omega_{ce}$, $\omega_{pe}$ and $\omega_{pi}$ are the ion cyclotron, the electron cyclotron, the electron plasma and the ion plasma frequencies   respectively. In this situation, the electrons are strongly magnetized while the ions are not. The beam width in the direction of propagation is chosen to be of the order of a few electron skin depths, $c/\omega_{pe}$, where $c$ is the velocity of light in vacuum.  The beam pulse, which acts as a moving current source for the electromagnetic excitations 
is considered to be rigid and free streaming with no loss in its energy as it propagates through the plasma. Its typical charge density is $ ~\sim 1.15n_0e$ where $n_0$ is the equilibrium plasma density and $e$ is the electronic charge.
A rectangular area of  $ 1500 c/\omega_{pe} \times 100 c/\omega_{pe} $)   in the $ x-y $ plane has been chosen for the simulations with $n_0=3.19\times 10^{20} cm^{-3}$ \cite{Kumar_2019}.
The spatial resolution chosen in the simulation is $10$ cells per electron skin depth ($c/\omega_{pe}$) with $64$ particles per cell for each species corresponding to a grid size of $\Delta x = 0.1 c/\omega_{pe}$ and the temporal resolution is given by the time step $ \Delta t = 0.0707 \omega_{pe}^{-1} $. A small initial noise in the form of a thermal velocity, $v_{the}=0.000442$ (corresponding to $0.1 eV$), is added to each species. The boundary conditions for the electromagnetic fields and particles are periodic in the $\hat{y}$  direction and absorbing in the $\hat{x}$ direction.  The value of the magnetic field is taken to be $B_{0} =2.5 m_ec\omega_{pe}/e$ where $m_e$ is the electronic mass. 

\section{Results}
\label{results}
\subsection{Electrostatic precursors} 
\label{ESP}
Prior to investigating electromagnetic precursors we carried out a series of simulations in the absence of the background magnetic field ($B=0$) in order to benchmark our code and to also examine the earlier fluid and molecular dynamic results of electrostatic precursors in the framework of a PIC simulation. The travelling pulsed beam now acts as a moving charge source and can give rise to electrostatic perturbations in the plasma and launch ion acoustic waves. To test the earlier fluid results we examine two cases, namely when (i) $V_b < V_{cs}$ and (ii) $V_b > V_{cs}$
where $V_b$ is the beam velocity and $V_{cs} = v_{the}/\sqrt{m_i/m_e}$ is the ion acoustic speed with $v_{the}$ being the electron thermal velocity. For our simulations we have chosen $V_{cs} = 0.1c$ and (i) $V_b=0.06c$ and (ii)$V_b=0.11c$  so that the two cases correspond to $M=0.6$ and $M=1.1$ respectively, where $M=V_b / V_{cs}$ is the Mach number.

\begin{figure}[h]
\center
                \includegraphics[width=0.9\textwidth]{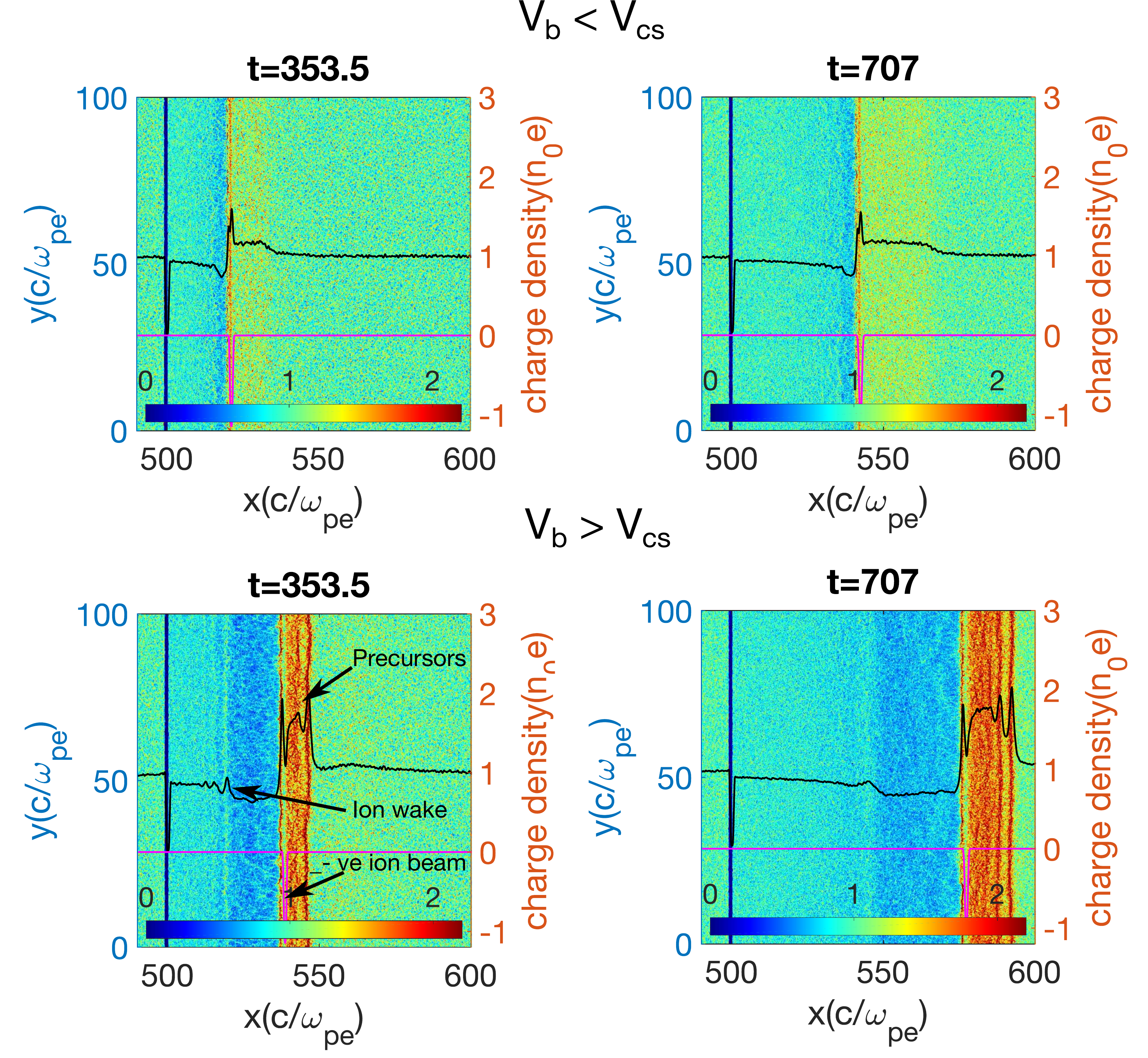} 
               
            \caption{Two dimensional snapshots of the electron density at t=353.5 and t=707 for subsonic ($V_b<V_{cs}$) and super-sonic ($V_b>V_{cs}$) propagation velocities of the beam. The line plots superposed on the snapshots are the $y$ averaged density values providing a one dimensional profile of the perturbations as a function of $x$. }  
                 \label{fig2}
                 
         \end{figure} 

Fig.~(\ref{fig2}) illustrates the simulation results for these two cases in the form of 2d snapshots of the electron density at $t=353.5$ and $t=700.0$. Superposed on the snapshots we have plotted the $y$ averaged density values (line plots) as a function of $x$. From the figures it is evident that for $V_b < V_{cs}$ one only 
gets wake fields excited behind the source whereas for $V_b>V_{cs}$ a series of precursor pulses are generated ahead of the traveling source. These precursor pulses are formed in front of the source due to the rapid piling up of the density at the expense of a density depression behind the source. The pile-up gives rise to the formation of a series of solitons which move faster than the source speed of $M=1.1$. We have independently checked the solitonic nature of these pulses by checking the constancy of the product of their amplitudes with the square of their widths - a procedure that will be described in more detail when we discuss the electromagnetic solitons.  
\begin{figure}[H]
\center
                \includegraphics[width=0.95\textwidth]{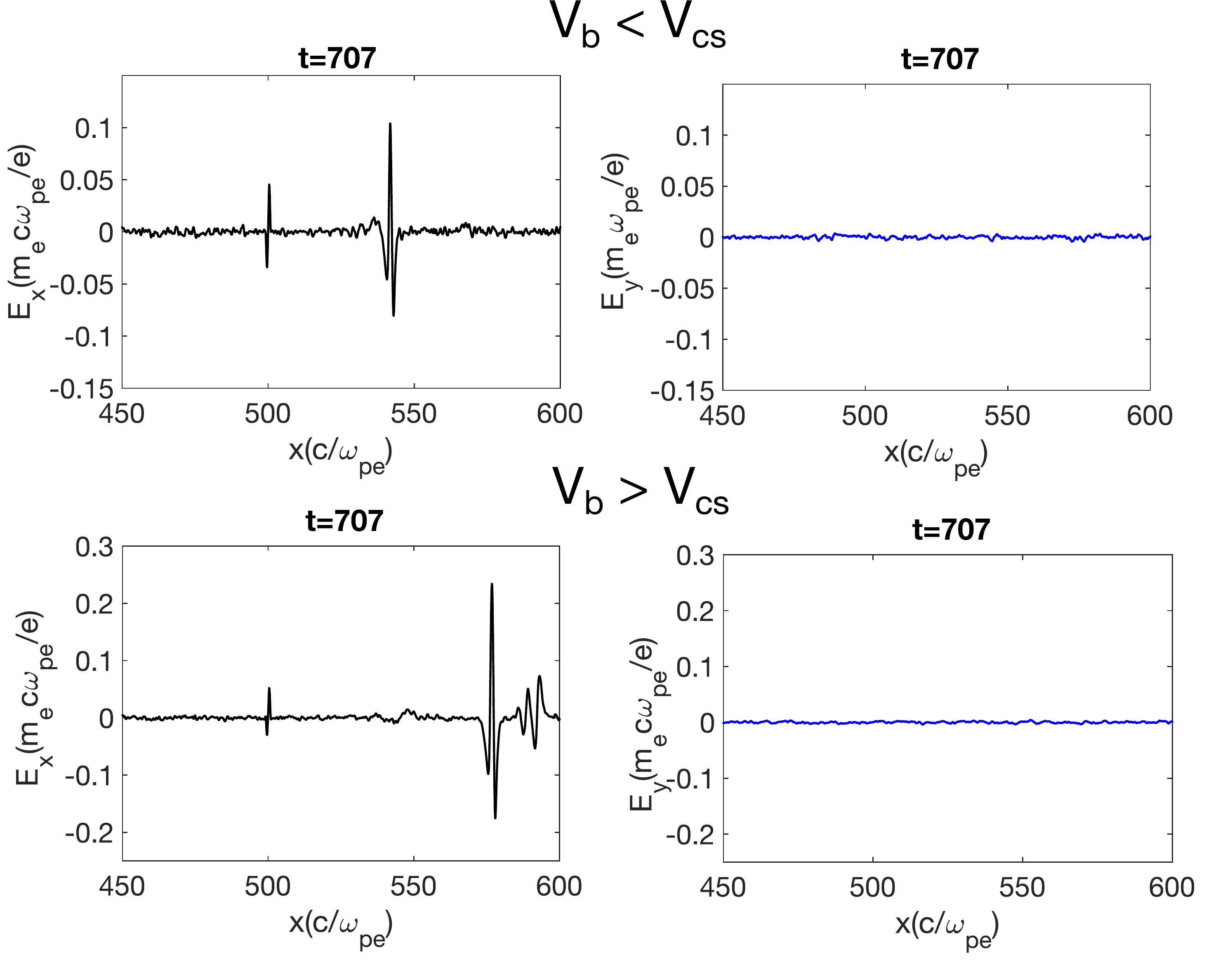} 
            \caption{Plots of the electric fields in the $x$ and $y$ directions for the $V_b <V_{cs}$ and $V_b>V_{cs}$ cases. }  
                 \label{fig3}
                 
         \end{figure} 
Fig.~(\ref{fig3}) shows the plots of the electric fields in the $x$ and $y$ directions. The absence of any $E_y$ further establishes the electrostatic nature of these excitations. 
Our electrostatic PIC simulation results are thus in accord with the earlier fluid and MD simulation results and not only serve as a good benchmark of our code but also provide a first-principles confirmation of this phenomenon at a particulate level. 

\subsection{Electromagnetic precursors}
\label{EMP}
\begin{figure}[h]
\center
                \includegraphics[width=\textwidth]{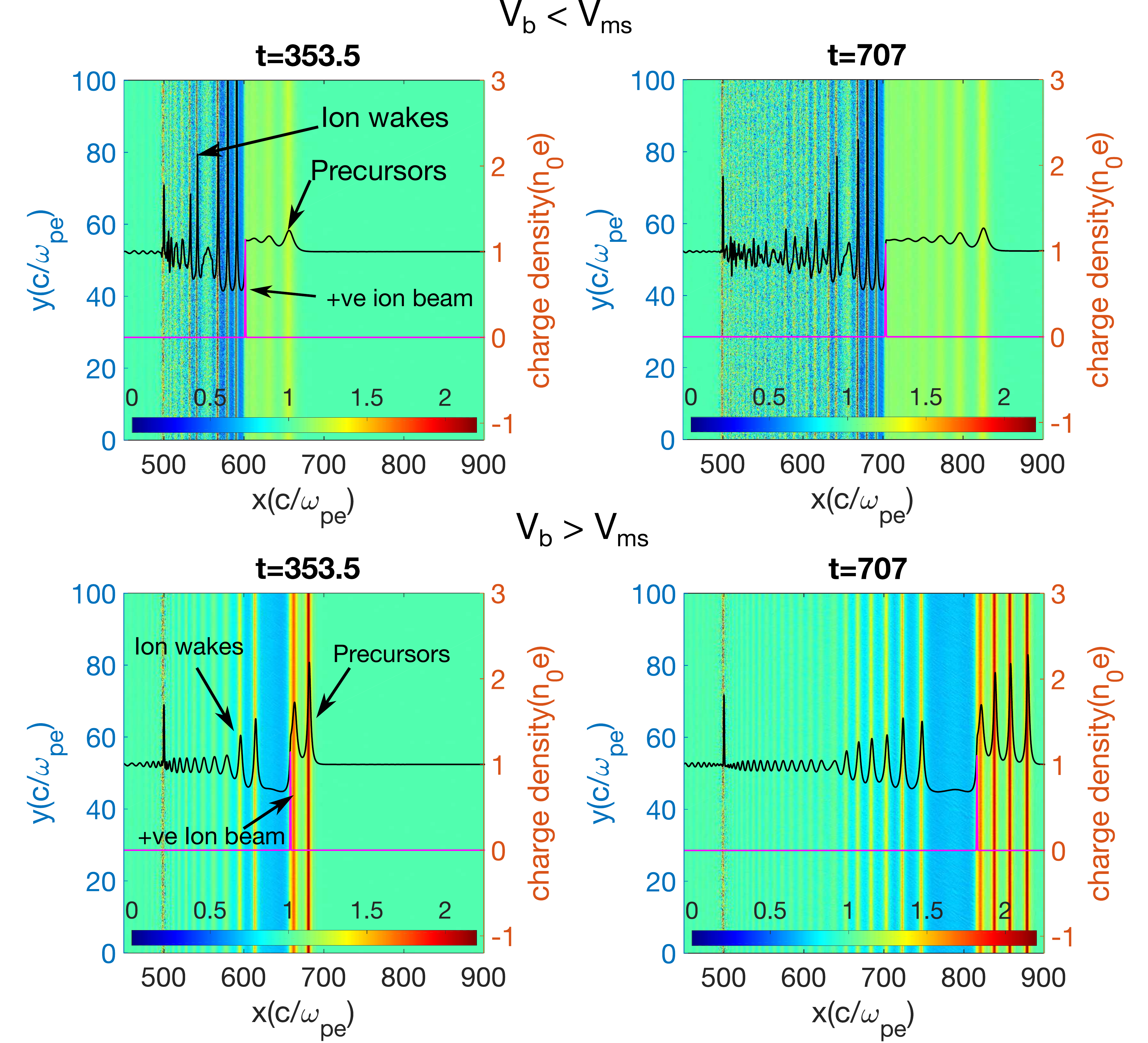} 
               
            \caption{Two dimensional snapshots of the electron density at t=353.5 and t=707 for ($V_b<V_{ms}$) and ($V_b>V_{ms}$) propagation velocities of the beam. The line plots superposed on the snapshots are the $y$ averaged density values providing a one dimensional profile of the perturbations as a function of $x$. }  
                 \label{fig4}
                 
         \end{figure}

We now discuss our simulation results carried out in the presence of an ambient magnetic field with the 
short pulse beam source traveling perpendicular to the field (as shown schematically in Fig.~\ref{fig1}).
We discuss two cases, namely, (i) $V_b < V_{ms}$ and (ii) $V_b > V_{ms}$, where $V_{ms}$ is the phase velocity of linear magneto-sonic waves and which for our cold plasma case is close to the Alfven velocity $V_A=(m_e/m_i)^{1/2}\omega_{ce}/\omega_{pe}= 0.5c$ for our choice of parameters. For our simulations we have chosen (i) $V_b=0.3c$ and (ii) $V_b= 0.505$ so that the two cases correspond to $M=0.6$ and $M=1.01$ respectively. Here $M= V_b/V_{ms}$ is the Mach number. Figure~\ref{fig4} illustrates the simulation results for these two cases where we have again superposed the $y$ averaged density values as line plots on the 2d density snapshots.

\begin{figure}
\center
                \includegraphics[width=\textwidth]{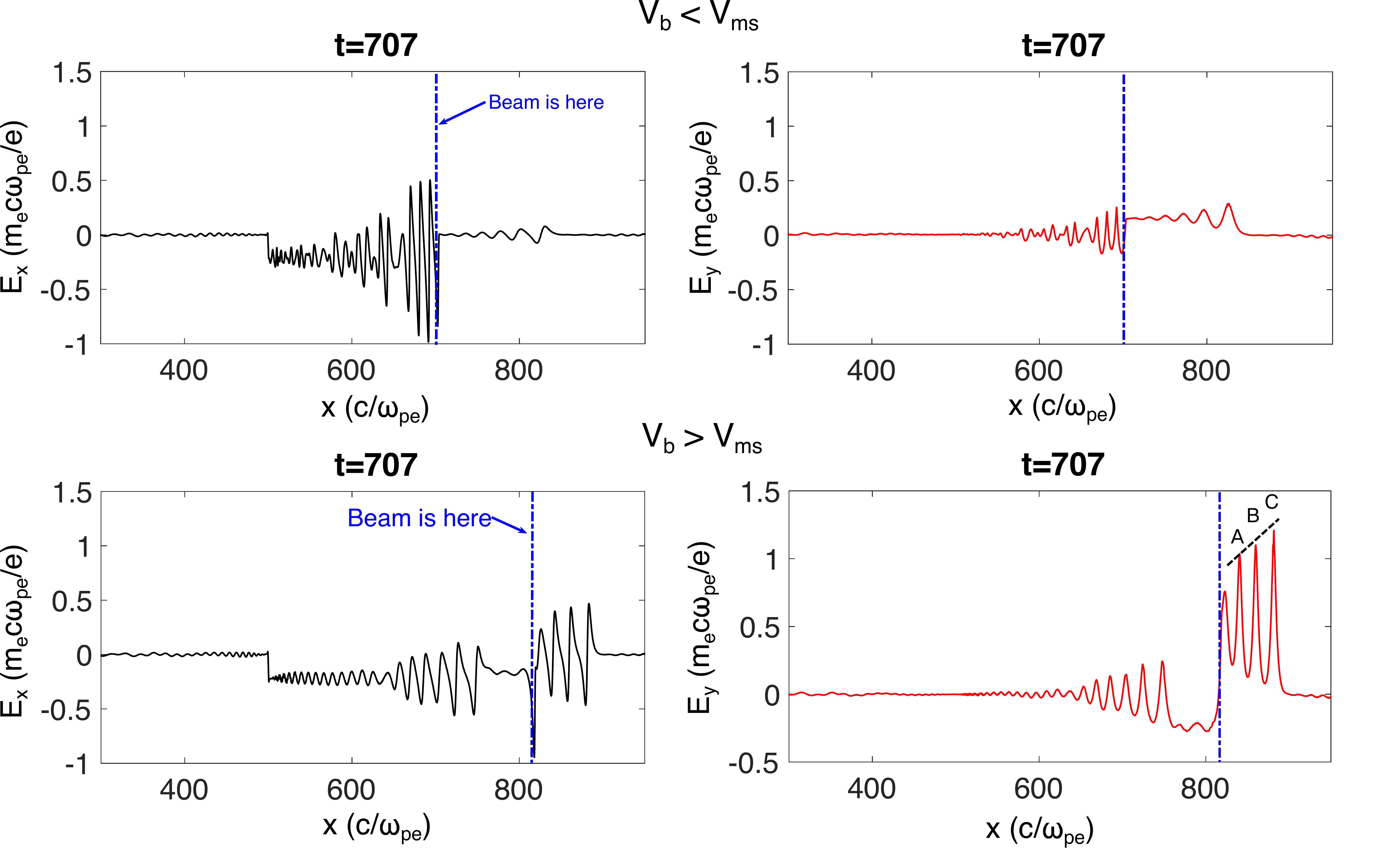} 
            \caption{Plots of $E_x$ and $E_y$ for the $V_b <V_{ms}$ and $V_b>V_{ms}$ cases.}  
                 \label{fig5}
                 
         \end{figure} 
                 

 Fig.~(\ref{fig5}) displays the electric field components of the wake-fields and precursors. Unlike the electrostatic case, we now find that the excitations have both an $E_x$ as well as an $E_y$ component thereby establishing their electromagnetic nature. These are, in fact, nonlinear magneto-sonic waves of the fast kind with the precursors taking the form of magneto-sonic solitons. In the present case the solitons are found to propagate with a Mach speed of around $M=1.13$ with the taller solitons moving faster than the smaller amplitude ones. In fact, their peaks lie on an ascending straight line as shown in Fig.~(\ref{fig5}) which is a characteristic of soliton solutions of the Korteweg de Vries (KdV) equation \cite{Zabusky1965}. Another signature of a KdV type soliton is the constancy of the product of its amplitude (a) with the square of their width (L). To further establish the solitonic nature of the precursors we have measured the amplitudes and widths of a number of them and checked the constancy of $aL^2$. Our findings are given in Table. I.

\begin{center}
{\bf{TABLE I}} \\

\vspace{0.2in}
\begin{tabular}{c c c c c c c c c c c c c  ll}
\hline
\hline
  &$Soliton $ \hspace{0.12in}  &$t (\omega_{pe}^{-1}) $ \hspace{0.12in}   & $a (\delta n_i/n_0)$  \hspace{0.12in}        &$L(c/\omega_{pe})$ \hspace{0.12in}        &$aL^2$  \\   
 \hline
 
  & A    \hspace{0.12in}      & 700\hspace{0.12in}     & 0.5325  \hspace{0.12in}      &4.1 \hspace{0.12in}    &   8.95  \\

  \hline

  & B    \hspace{0.12in}      & 700 \hspace{0.12in}     & 0.5865 \hspace{0.12in}      &3.9 \hspace{0.12in}    &   8.92   \\
  \hline

  & C    \hspace{0.12in}      & 700 \hspace{0.12in}     & 0.6605 \hspace{0.12in}      &3.6 \hspace{0.12in}    &   8.56 \\

\hline
\end{tabular}
  \end{center} 

As can be seen from the table, the parameter $ aL^2 $ changes very little in simulations and essentially remains constant 
for the observed precursor magnetosonic KdV solitons $A$, $B$ and $C$ of Fig.~(\ref{fig5}) . The percentage variation in $a$ is $24\%$ and $22\%$ for
$L^2$ but $aL^2$ varies only by $4.3\%$ in the data shown above.\\

We also notice from Fig.~(\ref{fig4}) that precursors are generated even when $V_b<V_{ms}$ but they are considerably weaker than the $V_b > V_{ms}$ case. This is distinctly different from the electrostatic case where for $V_b < V_{cs}$ no precursors were generated at all and therefore is suggestive of a fundamental difference in the excitation mechanism of the two kinds of solitons. Electrostatic solitons are born in front of the source pulse due to a rapid pile-up of the density when its velocity exceeds the phase velocity of the ion acoustic wave. The electromagnetic solitons, on the other hand, are primarily created in the wake region when the wakefields become nonlinear and can then overtake the source pulse due to their faster velocity when $V_b < V_{ms}$.   \\

Magneto-sonic solitons have been extensively studied in the past both analytically as well as in various numerical simulation studies. Most recently they have been observed in PIC simulations in the wake region of an intense laser pulse passing through a plasma \cite{Kumar_2019}. However, we believe they have never been studied or seen in the context of precursor pulses and our present results are the first to establish their existence ahead of a charged source traveling in a magnetized plasma.

\subsection{Pinned solitons}
\label{PIN}
\begin{figure}[H]
\center
                \includegraphics[width=0.95\textwidth]{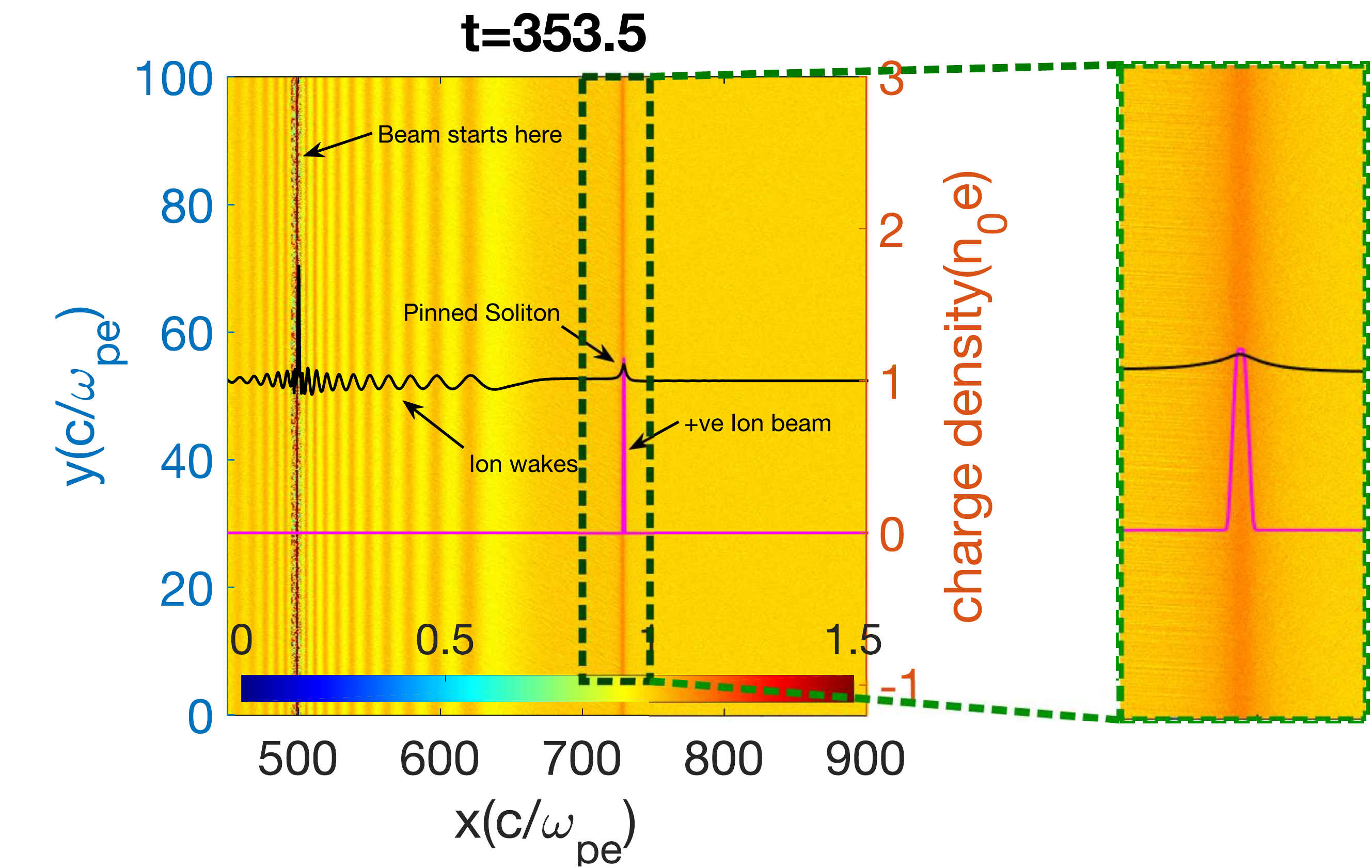} 
            \caption{The excitation of a pinned soliton when the beam travels at $V_b=0.85$ corresponding to $M=1.7$. The inset shows a magnified view of the structure of the soliton that envelops the source and travels at the same speed as the source. }  
                 \label{fig6}
                 
         \end{figure} 
         
\begin{figure}[H]
\center
                \includegraphics[width=0.95\textwidth]{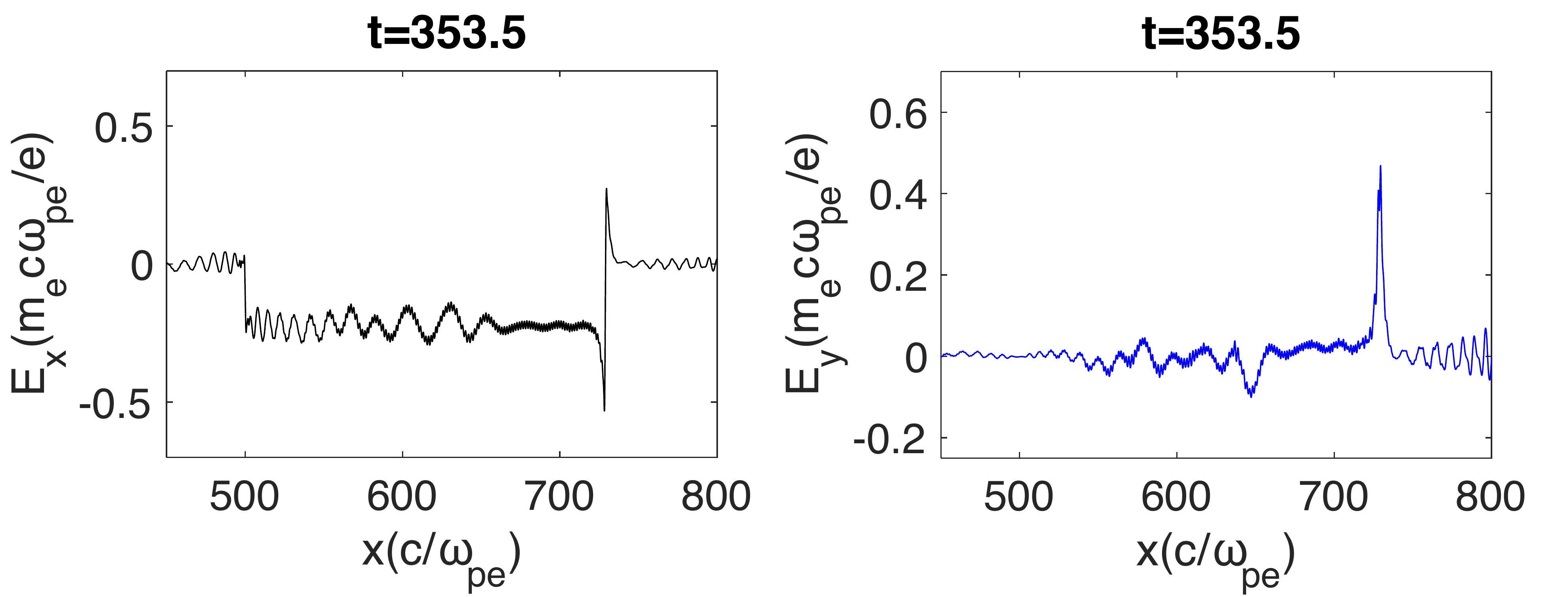} 
            \caption{Plots for $E_x$ and $E_y$ for Pinned soliton case where beam is moving with 0.85c (M=1.7). It is evident from the figure that the pinned solitons are also electromagnetic, magnetosonic solitons.}  
                 \label{fig7}
                 
         \end{figure}

As discussed in the previous section, magneto-sonic solitons are of the KdV type and propagate with velocities that have $M>1$. However as is well known \cite{davidson}, propagating KdV solitons only exist in a limited range of Mach numbers e.g. ($1<M<1.6$). So what happens when the source moves at a velocity that is beyond this range. To explore this question we have carried out simulations at $M=1.7$ and the results are shown in Fig~\ref{fig6}. One notices that there are no 
precursor solitons ahead of the moving source but an envelope structure develops around the source which travels at the same speed as the source. Such a structure known as a `pinned' soliton has been predicted theoretically for a 
driven KdV equation \cite{Wu1987,Jun-Xiao2009} but has not been captured in any PIC simulation studies. The structure is a few electron skin depths wide and has the characteristic bell shape form of a soliton. The longitudinal and  transverse electric field associated with the pinned soliton is shown in Fig.~\ref{fig7} establishing once more that these are electromagnetic structures. 

\section{Two dimensional sources}
\label{2d}
\begin{figure}[H]
\center
                \includegraphics[width=\textwidth]{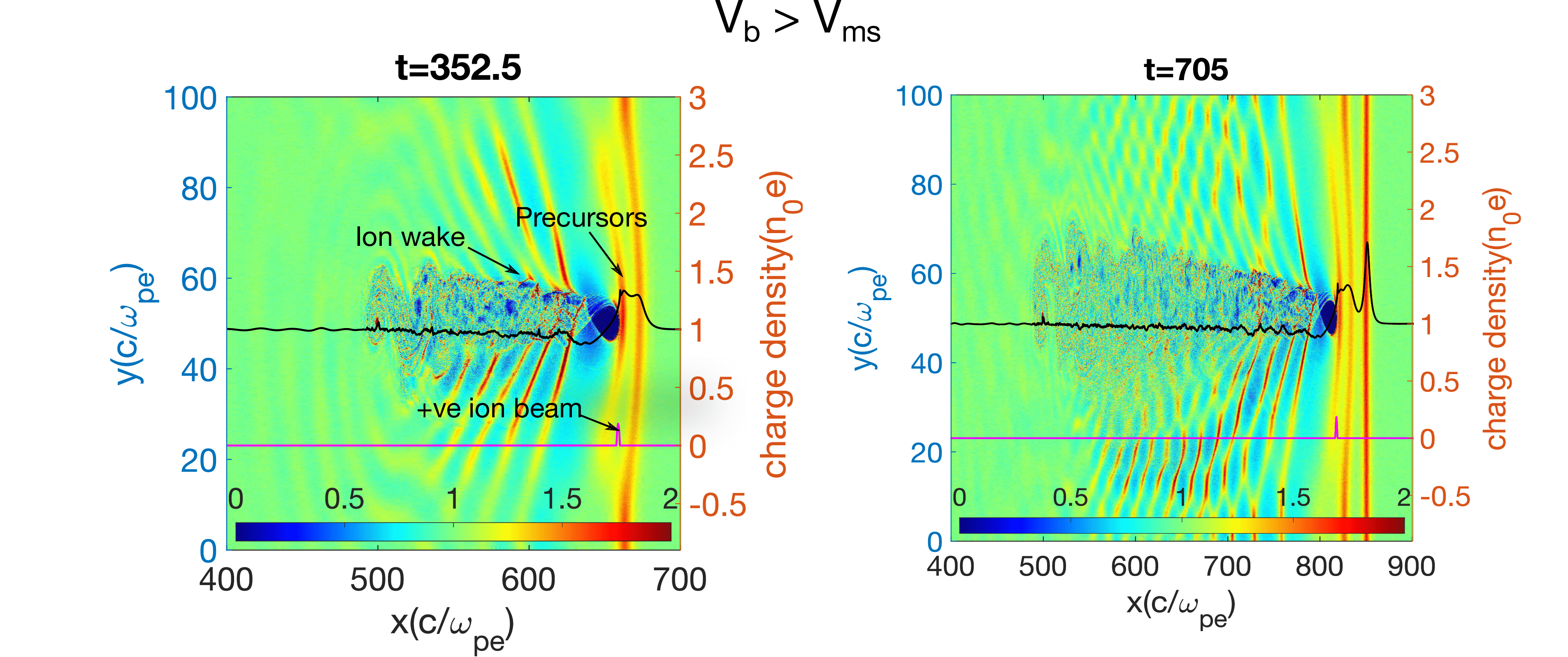}
            \caption{Typical snapshots of the electron density for a circular charged source moving at a Mach speed of 1.01 in the magnetized plasma.}  
                 \label{fig8}
                 
         \end{figure}

Our simulation results, presented in the preceding sections, were based on a simplification of the source term by assuming it to be of a one dimensional nature. This helped us to establish in principle the existence of the precursor and pinned solitons and to appropriately compare them to past fluid simulation results. However, in an actual physical situation, the moving charged object will be of a three dimensional nature and the size and shape of the object will likely influence the nature of the excitations. While a detailed study based on 3d-3V simulations is presently in progress and will be reported by us later, we present here a preliminary result obtained with a two dimensional circular source term that illustrates such an effect. Fig. \ref{fig8} shows a typical snapshot of the electron density for a circular source (of radius $r=2 c/\omega_{pe}$) moving at $M=1.01$ across the magnetic field in the plasma. The basic results remain the same, namely, the excitation of wake structures in the downstream region and precursor pulses in the upstream region.
The only noticeable difference from the 1d simulation is in the shape of the wakes and precursors which are now curved in nature. This means that the presence of the wakes and precursors will not be uniformly felt in the $y$ direction due to the curvature effect.   

\section{Discussion}
\label{disc}
To summarize, we have carried out detailed PIC simulations to study the phenomenon of precursor excitations of nonlinear pulses ahead of a charged source moving in a plasma. Our particular focus has been on the emission of electromagnetic pulses which have hitherto not been studied in this context. Our simulations establish for the first time the existence of such precursors in the form of magneto-sonic solitons that can either travel faster than the source or appear as stationary structures pinned to the source. Pinned solitons appear when the source speed is at a Mach number that is higher than about 1.6.  As a benchmark exercise, we have also confirmed the existence of electrostatic precursors in the form of ion acoustic solitons when the simulations were done in the absence of a magnetic field. These electrostatic results agree well with previously obtained fluid and molecular dynamic simulations. A comparison between the electrostatic and electromagnetic precursors reveals some interesting differences. The electrostatic wake fields are quite weak for low Mach numbers in comparison to those for the electromagnetic runs. Also while electrostatic precursors in the form of ion acoustic solitons occur only when the beam speed is supersonic ($V_b>V_{cs}$), the (electromagnetic) magnetosonic solitons appear even when $V_b<V_{ms}$. The difference can be traced to the basic mechanism for the creation of the solitons. The ion-acoustic solitons arise from the pile-up of the density in front of the charged source due to a balance between the nonlinear steepening and thermal broadening effects. Once formed they detach from the source and move away as their speed is faster than the source. 
The electromagnetic solitons on the other hand are not dependent on the density pile-up and are created from the large amplitude wake fields that arise in the downstream region. If the source is moving at a sub-magnetosonic speed then the solitons can easily overtake the source and move ahead of it. The wake field excitations get larger as the source speed approaches or slightly exceeds the magnetosonic speed and the resultant solitons then acquire a higher speed and again move ahead of the source. If the source speed exceeds a certain limit (in this case $M=1.6$) then
one does not get any precursors but the soliton generated from the wake sticks to the source as a pinned soliton.
Our simulations also reveal that the strength of the electromagnetic wake fields not only depends on the Mach number but also on the sign of the electrical charge of the source. Thus a positive ion beam source creates larger and sharper wakes than a corresponding negative ion beam of the same strength. Correspondingly the precursors created by the positive ion beam are sharper and larger in magnitude. These and other features of the wakes associated with the characteristics of the beam are currently under further investigation.\\

We now briefly discuss the possible natural occurrences of such solitons and also their potential practical applications. Flowing plasmas interacting with obstacles or charged objects moving through a plasma are common occurrences in nature. The most common example is the streaming solar wind plasma impinging on the earth and giving rise to the magnetosphere and a host of wave activity. From the perspective of our present simulations, one could view the earth as moving through the solar wind in the latter's frame of reference. The solar wind has components that can be supersonic or even super-Alfvenic. One can thus expect precursor pulses to be generated in the vicinity of the earth's bow shock and travel in the upstream (sun ward) direction. There have indeed been several satellite observations of upstream waves that range in frequency from mHz to tens of kHz \cite{Burgess1997} and comprise of both electrostatic and electromagnetic disturbances. Some of these structures have also been identified with electrostatic solitons and occasionally with electromagnetic solitons that correspond to magnetic holes or magnetosonic solitons. Our simulation results provide a new paradigm for the interpretation of these wave structures by  appropriately correlating them with the solar wind flow dynamics in these regions. Another area where such precursors might exist is in the ionosphere due to the interaction of objects like artificial satellites and space debris moving through the ambient plasma in this region. The objects get naturally  charged by collecting electrons and ions from the plasma and other mechanisms like photoionization and can therefore give rise to precursors as they travel. Such precursors traveling ahead of the objects can be detected from the ground using radar scattering for example, and could prove useful in tracking dangerous objects like small scale space debris that can damage live satellites \cite{Sen2015}. Electromagnetic solitons would be particularly useful in this regard as their spatial extent of a few electron skin depths would have a large footprint in the ionosphere and therefore be easy to remotely detect. 
\section{Acknowledgements}
\label{ack}
A.S. is thankful to the Indian National Science Academy
(INSA) for their support under the INSA Senior Scientist Fellowship
scheme and to AOARD for their research grant FA2386-18-1-4022. AK  would like to acknowledge the OSIRIS
Consortium, consisting of UCLA and IST(Lisbon, Portugal) for providing access to the OSIRIS4.0 framework. Work supported by NSF ACI-1339893. 
\section{References}

\bibliographystyle{unsrt}  

\bibliography{magnetosonic}

\begin{thebibliography}{10}

\bibitem{Sommerfeld1914}
A.~Sommerfeld.
\newblock Uber die fortpflanzung des lichtes in disperdierenden medien.
\newblock {\em Ann. Phys.}, 44:177 – 202, 1914.

\bibitem{Brillouin1914}
L.~Brillouin.
\newblock Uber die fortpflanzung des licht in disperdierenden medien.
\newblock {\em Ann. Phys.}, 44:204 – 240, 1914.

\bibitem{Brillouin1960}
L.~Brillouin.
\newblock {\em Wave Propagation and Group Velocity}.
\newblock Academic Press, New York, 1960.

\bibitem{Choi2004}
Seung-Ho Choi and Ulf \"Osterberg.
\newblock Observation of optical precursors in water.
\newblock {\em Phys. Rev. Lett.}, 92:193903, May 2004.

\bibitem{Albanese1989}
Richard Albanese, John Penn, and Richard Medina.
\newblock Short-rise-time microwave pulse propagation through dispersive
  biological media.
\newblock {\em J. Opt. Soc. Am. A}, 6(9):1441--1446, Sep 1989.

\bibitem{Jeong2006}
Heejeong Jeong, Andrew M.~C. Dawes, and Daniel~J. Gauthier.
\newblock Direct observation of optical precursors in a region of anomalous
  dispersion.
\newblock {\em Phys. Rev. Lett.}, 96:143901, Apr 2006.

\bibitem{Pleshko1969}
Peter Pleshko and Istv\'an Pal\'ocz.
\newblock Experimental observation of sommerfeld and brillouin precursors in
  the microwave domain.
\newblock {\em Phys. Rev. Lett.}, 22:1201--1204, Jun 1969.

\bibitem{Buckley1979}
R.~Buckley and C.~A.~P. Sammut.
\newblock Wave packets in dispersive media: an approximation encompassing both
  main signal and precursors.
\newblock {\em Journal of Plasma Physics}, 22(3):563–570, 1979.

\bibitem{Cartwright2009}
N.~Cartwright and K.E. Oughstun.
\newblock Ultrawideband pulse propagation through a homogeneous, isotropic,
  lossy plasma.
\newblock {\em Radio Science}, 44, 2009.

\bibitem{Ertekin1984}
R.~C. Ertekin, W.~C. Webster, and J.~V. Wehausen.
\newblock Ship-generated solitons.
\newblock {\em Proc. 15th Symp. Naval Hydrodyn., Hamburg.}, 1984.

\bibitem{Akylas1984}
T.~R. Akylas.
\newblock On the excitation of long nonlinear water waves by a moving pressure
  distribution.
\newblock {\em J. Fluid Mech.}, 141:455--466, 1984.

\bibitem{Lee1989}
Seung-Joon Lee, George~T. Yates, and T.~Yaotsu Wu.
\newblock Experiments and analyses of upstream-advancing solitary waves
  generated by moving disturbances.
\newblock {\em Journal of Fluid Mechanics}, 199:569--593, 2 1989.

\bibitem{Sen2015}
Abhijit Sen, Sanat Tiwari, Sanjay Mishra, and Predhiman Kaw.
\newblock Nonlinear wave excitations by orbiting charged space debris objects.
\newblock {\em Advances in Space Research}, 56(3):429 -- 435, 2015.

\bibitem{Sanat_prec}
Sanat~Kumar Tiwari and Abhijit Sen.
\newblock Wakes and precursor soliton excitations by a moving charged object in
  a plasma.
\newblock {\em Physics of Plasmas}, 23:022301, 2016.

\bibitem{Sanat_md2016}
Sanat~Kumar Tiwari and Abhijit Sen.
\newblock Fore-wake excitations from moving charged objects in a complex
  plasma.
\newblock {\em Physics of Plasmas}, 23(10):100705, 2016.

\bibitem{Surabhi2016}
S.~{Jaiswal}, P.~{Bandyopadhyay}, and A.~{Sen}.
\newblock {Experimental observation of precursor solitons in a flowing complex
  plasma}.
\newblock {\em Phys. Rev. E}, 93:041201(R), 2016.

\bibitem{Garima2019}
Garima Arora, P.~Bandyopadhyay, M.~G. Hariprasad, and A.~Sen.
\newblock Effect of size and shape of a moving charged object on the
  propagation characteristics of precursor solitons.
\newblock {\em Physics of Plasmas}, 26(9):093701, 2019.

\bibitem{hemker}
R.~G. {Hemker}.
\newblock {Particle-In-Cell Modeling of Plasma-Based Accelerators in Two and
  Three Dimensions}.
\newblock {\em Thesis, University of California, Los Angeles}, 2000.

\bibitem{Fonseca2002}
R.~A. Fonseca, L.~O. Silva, F.~S. Tsung, V.~K. Decyk, W.~Lu, C.~Ren, W.~B.
  Mori, S.~Deng, S.~Lee, T.~Katsouleas, and J.~C. Adam.
\newblock {\em OSIRIS: A Three-Dimensional, Fully Relativistic Particle in Cell
  Code for Modeling Plasma Based Accelerators}, pages 342--351.
\newblock Springer Berlin Heidelberg, Berlin, Heidelberg, 2002.

\bibitem{osiris}
R~A Fonseca, S~F Martins, L~O Silva, J~W Tonge, F~S Tsung, and W~B Mori.
\newblock One-to-one direct modeling of experiments and astrophysical
  scenarios: pushing the envelope on kinetic plasma simulations.
\newblock {\em Plasma Physics and Controlled Fusion}, 50(12):124034, 2008.

\bibitem{Kumar_2019}
Atul Kumar, Chandrasekhar Shukla, Deepa Verma, Amita Das, and Predhiman Kaw.
\newblock Excitation of {KdV} magnetosonic solitons in plasma in the presence
  of an external magnetic field.
\newblock {\em Plasma Physics and Controlled Fusion}, 61(6):065009, may 2019.

\bibitem{Zabusky1965}
N.~J. Zabusky and M.~D. Kruskal.
\newblock Interaction of ``solitons'' in a collisionless plasma and the
  recurrence of initial states.
\newblock {\em Phys. Rev. Lett.}, 15:240--243, 1965.

\bibitem{davidson}
R.C. Davidson.
\newblock {\em Methods in Nonlinear Plasma Theory}.
\newblock Academic Press, New York, 1972.

\bibitem{Wu1987}
T.~Y. Wu.
\newblock Generation of upstream advancing solitons by moving disturbances.
\newblock {\em J. Fluid Mech.}, 184:75--99, 1987.

\bibitem{Jun-Xiao2009}
Zhao Jun-Xiao and Guo Bo-Ling.
\newblock {Analytic solutions to forced KdV equation}.
\newblock {\em Commun. Theor. Phys. (Beijing, China)}, 52(2):279--283, 2009.

\bibitem{Burgess1997}
D.~Burgess.
\newblock What do we really know about upstream waves?
\newblock {\em Advances in Space Research}, 20:673--682, 1997.

\end{thebibliography}

\end{document}